\documentclass[draft,prb,twocolumn,showpacs,nobibnotes]{revtex4}

\usepackage{bm}
\usepackage[dvips,final]{graphics}

\begin{document}

\title{Violation of the Wiedemann-Franz Law in a Large-N Solution of the t-J Model} 

\author{A. Houghton}
\author{S. Lee}
\author{J. B. Marston} 
\affiliation{Department of Physics, Brown University, Providence, RI
02912-1843}

\date{\today}  

\begin{abstract}
We show that the Wiedemann-Franz law, which holds for Landau Fermi liquids, 
breaks down in a large-n treatment of the t-J model.  
The calculated ratio of the in-plane 
thermal and electrical conductivities agrees quantitatively with 
experiments on the normal state of the electron-doped Pr$_{2-x}$Ce$_x$CuO$_4$ ($x = 0.15$) 
cuprate superconductor. 
The violation of the Wiedemann-Franz law in the uniform phase contrasts with 
other properties of the phase that are Fermi liquid like.
\end{abstract}


\maketitle

A recent experiment that measured the electrical and thermal conductivities of
the copper-oxide superconductor Pr$_{2-x}$Ce$_x$CuO$_4$ in its normal state 
found striking deviations from the Wiedemann-Franz law\cite{hill}.  Simply
stated, the Wiedemann-Franz law says that fermion quasiparticles transport
both electrical and heat currents, with the ratio of the 
heat conductivity $\kappa$ to the electrical conductivity $\sigma$ given by:
\begin{equation}
{{\kappa}\over{\sigma T}} = {{\pi^2}\over{3}}~ \bigg{(} {{k_B}\over{e}} \bigg{)}^2
\end{equation}
Ordinary Landau Fermi liquids respect the Wiedemann-Franz law, so deviations from
it indicate the presence of non-Fermi liquid physics.  
The results of Ref.~\onlinecite{hill}
are therefore broadly consistent with other experimental 
evidence that points to non-Fermi 
liquid behavior in the cuprate phase diagram. 
 
One possible interpretation of the breakdown of the Wiedemann-Franz law 
is that the quasiparticle fractionalizes into separate carriers of charge
and spin.  To see what effects such fractionalization might have, 
consider first heat transport in a 
system of weakly-interacting electrons, which could be viewed as a crude approximation
to electrons in the highly overdoped region of the cuprate phase diagram.  
The electrons transport both charge and heat. The electrical conductivity is given by 
the Drude formula $\sigma = n e^2 \tau / m$ and the thermal conductivity $\kappa \propto T$ because
while each quasiparticle carries fixed charge $e$, it only carries an
energy of order the temperature.  Next consider a model for the undoped cuprates: the N\'eel 
ordered antiferromagnetic insulator with zero  electrical conductivity.
Phonons and spinwaves transport heat, and as both excitations are bosonic in character 
with linear dispersions at low energy, each contributes 
similarly, yielding a thermal conductivity $\kappa \propto T^3$.  

Leaving aside these ordinary states of matter, consider the case in which the spins in 
the insulator, instead of ordering, fractionalize into spinons with an extended 
Fermi surface and a non-zero density of states\cite{PWA,BZA}.  
Now spins transport heat in much the same way as charges do
in the non-interacting electron system, with $\kappa \propto T$.  The Lorenz ratio 
is infinite, and remains significantly larger than one upon doping with holes or electrons.    

Just this scenario is predicted in a large-n treatment of the t-J model on the 
square lattice.  We follow the approach of Refs.~\onlinecite{AM,MA,Ted}.    
Implementing the single-occupancy constraint by introducing slave-boson operators
$b_i$, the t-J model may be written: 
\begin{eqnarray}
H &=& -t~ \sum_{<i,j>}~ (c^{\dagger \alpha}_i b_i c_{j \alpha} b^\dagger_j + H.c.)
\nonumber \\
&+& J~ \sum_{<i,j>}~ (\vec{S}_i \cdot \vec{S}_j - \frac{1}{4} n_i n_j)
\nonumber \\
&+& {{1}\over{2}}~ \sum_{i \neq j}~ V(|\vec{r}_i - \vec{r}_j|)~ n_i n_j \ .
\label{tJ}
\end{eqnarray}
Number and spin operators are related to the electron creation and annihilation 
operators by $n_i \equiv c^{\dagger \alpha}_i c_{i \alpha}$ and
$\vec{S}_i \equiv (1/2) c^{\dagger \alpha}_i \vec{\sigma}_\alpha^\beta c_{i \beta}$
where there is an implicit sum over pairs of raised and lowered Greek indices.
The single-occupancy constraint is now holonomic, 
$b^\dagger_i b_i + c^{\dagger \alpha}_i c_{i \alpha} = 1$,
with the physical meaning that only a hole, or a single electron, may occupy each site
of the lattice.  (In the following 
we generally refer to hole doping with the understanding
that our calculations apply equally well to electron doped systems.) 
Included in Eq.~\ref{tJ} is the off-site Coulomb repulsion $V(r)$; however, in the uniform
and staggered flux phases discussed below it plays no role other than to contribute an 
additive constant to the total energy. 

Because spin-exchange involves no net flow of charge, electrical current 
only arises from the hopping term.  The continuity equation relates the time
rate of change of the occupancy on a given site to the lattice divergence of the 
current flowing into the site:
\begin{equation}
e * {{d n_j(t)}\over{dt}} = -i * e * [n_j,~ H] 
= - \sum_{\hat{e} = \hat{x}, \hat{y}} {{j^e_{j, j + \hat{e}} - j^e_{j - \hat{e}, j}}\over{a}}  
\end{equation}
where $a$ is the lattice spacing between copper atoms.  Thus
\begin{equation}
j^e_{j, j + \hat{e}} = -i e * t * a * (c^{\dagger \alpha}_{j} b_j 
c_{j + \hat{e} \alpha} b^\dagger_{j + \hat{e}} - H.c.)
\end{equation}
is the electric current flowing from site $j$ into site $j+\hat{e}$ along the link connecting
the two sites.  We emphasize that neither the spin-spin exchange interaction $J$ 
nor the Coulomb interaction $V(r)$ appear in the expression
for the electrical current.  This result, which is a direct consequence 
of the gauge invariance of the spin-exchange and Coulomb interactions 
in our microscopic calculation, contrasts with that obtained recently\cite{yang} 
within the more phenomenological ``d-density wave'' picture\cite{nayak}.  

The heat current $j^q_{j, j+\hat{e}}$ can be found by taking the time derivative of the 
Hamiltonian density:
\begin{equation}
{{d h_j(t)}\over{dt}} = 
- \sum_{\hat{e} = \hat{x}, \hat{y}} {{j^q_{j, j+\hat{e}} - j^q_{j - \hat{e}, j}}\over{a}}  
\end{equation}
where 
\begin{equation}
h_j = \sum_{i = j \pm \hat{x}, \hat{y}}~ (-t~ c^{\dagger \alpha}_i b_i c_{j \alpha} b^\dagger_j + H.c.)
+ J~ \vec{S}_i \cdot \vec{S}_j  
\label{ham-density}
\end{equation}
and the sum is only over the four sites $i$ that are nearest-neighbors of site $j$. 
We have dropped the $n_i n_j$ interaction terms.  These do not contribute to
the DC thermal conductivity because $\langle n_i \rangle$ remains unchanged in 
the presence of currents.  However, in contrast to the electrical current, 
the heat current has contributions both from hopping, and from spin-exchange:
\begin{eqnarray}
j^q_{j, j+\hat{e}} &=& t * a * [ c^{\dagger \alpha}_{j} b_j 
\partial_t (c_{j + \hat{e} \alpha} b^\dagger_{j + \hat{e}}) + H.c.]
\nonumber \\
&+& {{J * a}\over{2}} * [ c^{\dagger \alpha}_j (\partial_t c_{j + \hat{e} \alpha}) 
c^{\dagger \beta}_{j + \hat{e}} c_{j \beta} 
\nonumber \\
&+& c^{\dagger \alpha}_j c_{j + \hat{e} \alpha} 
(\partial_t c^{\dagger \beta}_{j + \hat{e}}) c_{j \beta} ]\ . 
\label{jq}
\end{eqnarray}

The model generalizes from the physical case of $n = 2$ (up and down spins) 
to arbitrary (even integer) values of $n$ by letting the Greek indices run over $1, \cdots, n$.
In the large-n limit the spin-spin interaction factorizes in the particle-hole channel.
Formally this factorization is implemented via a Hubbard-Stratonovich transformation 
within the functional integral approach. Complex-valued mean-fields along the bonds
are introduced:
\begin{equation}
\chi_{i j} = \frac{J}{n} \langle c^{\dagger \alpha}_i c_{j \alpha} \rangle\ .
\end{equation}
The $\chi_{ij}$ fields function as the order parameter, and as they are spin-rotation 
invariant, there is no possibility
of N\'eel or other spin order, and the mean-field Hamiltonian respects spin-rotational symmetry.
Furthermore, at sufficiently low temperatures the (holon) boson
fields condense\cite{condense}, 
and we may make the replacement $b_i = b^\dagger_i = \sqrt{x}$ where
$x$ is the hole density.  The mean-field Hamiltonian, which is exact in the 
$n \rightarrow \infty$ limit, then takes the form: 
\begin{equation}
H_{MF} = \sum_{<i,j>}~ [ {{n}\over{J}}~ |\chi_{i j}|^2 -
(t~ x + \chi_{i j}) (c^{\dagger \alpha}_j c_{i \alpha} + H.c.) ]\ .
\label{H_mf}
\end{equation}
For parameters appropriate to the cuprates, $t = 0.44$ eV (following
Hybertsen {et al.}\cite{hybertsen}) and $J = 0.13$ eV (obtained by
Singh {\it et al.}\cite{singh}),
the minimum energy configuration has $\chi_{ij}$ both real and constant 
when the doping exceeds $x > 0.12$.  There are no broken symmetries in this 
``uniform'' phase.  Upon suppressing dimerization with the addition of a
biquadratic spin exchange interaction\cite{MA,Vojta}, 
a staggered flux (SF) phase with counter-circulating
orbital currents\cite{AM,MA,Ted,john} occurs for $x < 0.12$, that is, in the
underdoped region of the phase diagram. (The biquadratic interaction which simultaneously
exchanges four fermions disappears in the 
physical $n = 2$ limit of up and down spins,
and thus does not alter the physics of the t-J model\cite{MA}.)
More realistic models include non-zero next- and third-nearest neighbor 
hopping amplitudes\cite{OKA} but these terms do not qualitatively affect 
the phase diagram or transport behavior.

The expressions for the two currents simplify in the large-n limit.  As the boson operators
may be replaced by the c-number $\sqrt{x}$, the electrical current becomes: 
\begin{equation}
j^e_{j, j+\hat{e}} = -i e * t * x * a * (c^{\dagger \alpha}_{j} c_{j + \hat{e} \alpha} - H.c.) \ .
\label{je-mf}
\end{equation}
Upon further replacing the fermion bilinear operator $c^{\dagger \alpha}_i c_{j \alpha}$ 
with ${{n}\over{J}}~ \chi$, the heat current also simplifies\cite{why}: 
\begin{equation}
j^q_{j, j+\hat{e}} = (t * x + \chi) * a * (c^{\dagger \alpha}_{j} \partial_t c_{j + \hat{e} \alpha} + H.c.)\ .
\label{jq-mf}
\end{equation}
As the heat current differs only by the $(t * x + \chi)$ prefactor from that of a 
non-interacting tight-binding system, in the low-temperature limit 
the Lorenz ratio for in-plane transport is simply:
\begin{equation}
{{\kappa}\over{\sigma T}} = {{\pi^2}\over{3}}~ \bigg{(} {{k_B}\over{e}} \bigg{)}^2 
* \bigg{(} {{t x + \chi}\over{t x}} \bigg{)}^2\ . 
\label{wf}
\end{equation}
Thus for any $\chi \neq 0$ the ratio differs from unity, indicating a breakdown of 
Fermi liquid theory.  Note that all details of the scattering mechanisms cancel out in
the ratio.  Direct calculation of the two conductivities in linear response shows that
the integrals over momentum have identical form. Only the frequency integrals differ; 
hence for static impurities in the weak-scattering limit 
the Lorenz ratio is given by Eq.~\ref{wf}.  We note that while the order
parameter $\chi$ is perturbed by the application of external electric fields
or thermal gradients, the perturbation in $\chi$ does not alter the DC response\cite{caroli}. 

In the SF phase the $\chi$-fields are complex numbers\cite{AM,MA,Ted}, 
and the prefactor $(t x + \chi)^2$ should be replaced by $|t x + \chi |^2$.  
Specifically, for $\chi = |\chi| \exp(i \theta)$, with phase $\theta$, 
the Lorenz ratio generalizes to\cite{in_prep}: 
\begin{equation}
{{L}\over{L_0}} =  
{{(t x)^2  + |\chi|^2 + 2 t x |\chi| \cos(\theta)}\over{(t x)^2}}  
\label{wf-sf}
\end{equation}
where $L \equiv \kappa / (\sigma T)$ 
and  $L_0 \equiv (\pi^2/3) (k_B /e)^2 \approx 2.45 \times 10^{-8} {\rm W \Omega K^{-2}}$ 
is the Lorenz number.

\begin{figure}[h,t]
\resizebox{8cm}{!}{\includegraphics{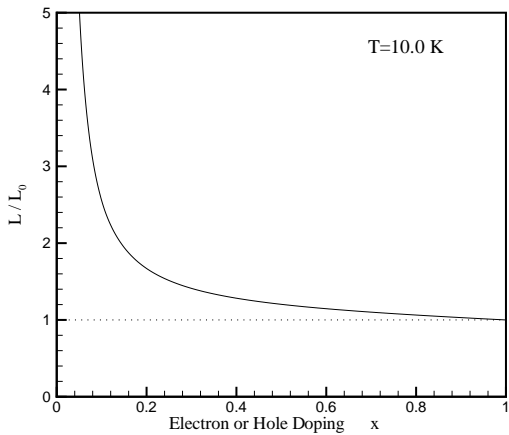}}
\caption{The Lorenz ratio (Eqs.~\ref{wf} and \ref{wf-sf}) as
a function of the doping for $t = 0.44$ eV, $J = 0.13$ eV.
The ratio approaches unity in the dilute limit, $x \rightarrow 1$.}
\label{fig:ratio-x}
\end{figure}

We turn now to a comparison of our predictions, Eqs.~\ref{wf} and \ref{wf-sf}, 
with experiments.  In Fig.~\ref{fig:ratio-x}
we plot the Lorenz ratio as a function of the doping.  As expected, the 
ratio diverges as the insulating limit $x \rightarrow 0$ is approached because 
the spinons transport only heat, not charge.  In the opposite limit of large doping 
$\chi \rightarrow 0$, and the ratio approaches unity. Landau Fermi liquid  
theory is recovered in the dilute limit of widely spaced electrons.  
We emphasize that the uniform phase with
$\chi_{ij}$ constant and real does not break any symmetries.  
It exhibits weak pseudogap behavior 
because, according to the mean-field equations, $|\chi|$ increases
slightly in size at low temperatures, which in turn increases 
the quasiparticle bandwidth (see Eq.~\ref{H_mf}) 
and decreases the density of states (DOS)\cite{chung}. 
For example, at a hole concentration of 
$x = 0.15$, $|\chi| = 0.024$ eV at $T = 500$ K rising
to $|\chi| = 0.026$ eV at zero temperature; consequently the DOS drops by 2\%.  
This contrasts with the strong pseudogap behavior of the SF phase which has a gap along 
most of the Fermi surface and which breaks time reversal invariance.  
In either phase, however, the fermionic quasiparticles are non-interacting in
the $n \rightarrow \infty$ limit and hence behave as long-lived Landau quasiparticles 
such as those found in ordinary Fermi liquids.  

We note that the Wiedemann-Franz law is strongly violated in s-wave superconducting
states because while Cooper pairs carry charge, the condensate has no entropy. 
In a d-wave superconductor, quasiparticle excitations at the nodes 
result in a modified Wiedemann-Franz law\cite{durst}.  
The violation that we find occurs in the normal state, and is a consequence of 
the spin-charge separation inherent in the large-n solution of the t-J model,
and not of any incipient superconducting tendencies. 

\begin{figure}[h,t]
\resizebox{8cm}{!}{\includegraphics{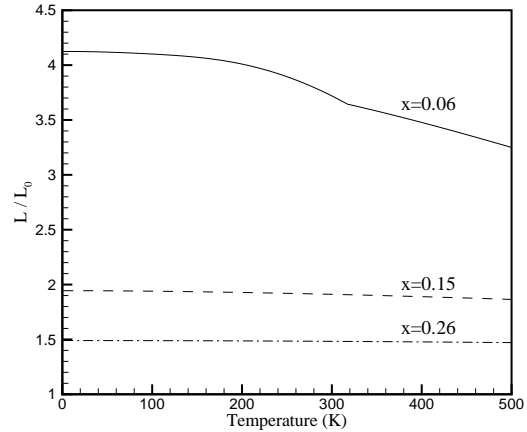}}
\caption{The Lorenz ratio (Eqs.~\ref{wf} and \ref{wf-sf}) as a function of temperature
at dopings $x = 0.06$, $0.15$, and $0.26$ 
for $t = 0.44$ eV, $J = 0.13$ eV.  In the underdoped $x = 0.06$ case
there is a transition to the staggered flux phase at a temperature of
approximately $300$ K.} 
\label{fig:ratio-T}
\end{figure}

In Fig.~\ref{fig:ratio-T} we plot the temperature dependence of the 
Lorenz ratio for three dopings at which transport experiments have been 
conducted: $x = 0.06$, $0.15$, and $0.26$. For a single crystal of the
La$_{2-x}$Sr$_x$CuO$_4$ material with hole doping $x = 0.06$, the resistivity
was measured upon suppressing the superconductivity by application of a 
18 T magnetic field along the c-axis\cite{takeya}.  The thermal conductivity
was, however, measured in the superconducting state, so it is not possible to 
extract a real Lorenz ratio.  Nevertheless it is intriguing that 
$L / L_0 \approx 5$ at low temperatures, based on the numbers appearing in the
inset to Fig. 3 of Ref.~\onlinecite{takeya}.  This compares 
reasonably well with the theoretical value of $4.1$ seen in Fig.~\ref{fig:ratio-T}.
    
At optimal doping, experimentally available magnetic fields can only eliminate
superconductivity in electron-doped compounds.  In Ref.~\onlinecite{hill} a 13 T field
was applied to Pr$_{2-x}$Ce$_x$CuO$_4$ at $x = 0.15$ to access the normal state. 
The measured ratio of $L / L_0 \approx 2$ found at temperatures
above $0.3$ K is again in reasonable quantitative agreement with 
the theoretical value of $1.95$.  At temperatures below $0.18$ K, however,
the experimentally determined ratio drops rapidly below one.  We have no explanation for
the observed behavior at the lowest temperatures\cite{si}.

Finally, in the highly overdoped regime Proust {\it et al.}\cite{proust} 
have studied the Tl$_2$Ba$_2$CuO$_{6+\delta}$ material at a hole concentration
of $x \approx 0.26$.  Superconductivity was suppressed in a 13 T field, and
$L / L_0 = 0.99 \pm 0.01$ in good agreement with the
Wiedemann-Franz law for Fermi liquids.  The theoretical value of the ratio is $1.5$.
We speculate that the persistence of non-Fermi liquid behavior at large doping
in the mean-field theory is an artifact of the large-n approximation.  Finite-n 
corrections could possibly restore Fermi liquid behavior in the overdoped region.  
At large doping $|\chi|$ is small compared to the effective hole hopping amplitude $t x$,
so fluctuations in $\chi$ 
may be expected to be relatively more important than at low doping.

In summary we have shown that the Wiedemann-Franz law is 
violated in a mean-field treatment of the t-J model. 
Our analysis, which holds for weak scattering, is exact in the 
$n \rightarrow \infty$ limit.  The Lorenz ratio is significantly larger than one
both in the uniform phase ($x > 0.12$) and in the SF phase ($x < 0.12$).  
The theoretical prediction is in reasonably good quantitative agreement
with existing experimental measurements on the cuprate materials.  

{\it Note added: } After this work was completed a paper by Kim and Carbotte (KC)
appeared\cite{kim} that examined the Wiedemann-Franz law within the context of
the phenomenological d-density wave picture.  There are several differences between 
their work and ours.  The main difference is that we study both the 
uniform phase which has no broken symmetries, and the SF or d-density phase with
time-reversal breaking counter-circulating currents.
We find that the Lorenz ratio is significantly larger than one in both phases.
Furthermore, at low temperatures KC find only small deviations 
from the Wiedemann-Franz law.  This is due in part to the fact that their d-density 
order parameter (the analog of our $\chi_{ij}$) was chosen to be purely imaginary 
(equivalent to setting $\theta = \pi/2$ in our Eq.~\ref{wf-sf}) and also because  
their kinetic energy is not rescaled by the slave-boson doping factor, $x$, as it
is in our microscopic analysis of the t-J model.  KC also find a 
large temperature variation in the Lorenz ratio for the case of
strong scattering because the quasiparticle lifetime has a strong frequency dependence.

We thank John Fj{\ae}restad and Louis Taillefer for helpful comments.
This work was supported in part by the NSF under grant No. DMR-9712391.


\begin{thebibliography}{}
\bibitem{condense} Strictly speaking bosons cannot condense in a purely
two-dimensional system at any non-zero temperature.  In practice, however, weak
interlayer tunneling stabilizes the condensate.  As the temperature scale for
condensation is determined primarily by the in-plane hopping amplitude $t$, 
and since $t \gg k_B T$, we
assume that the bosons remain condensed at the temperatures relevant for the
experiments we consider here.
\bibitem{why} It is straightforward to verify that the heat 
current-current correlation function is the same in the large-n limit 
for either form of the current operator, Eq.~\ref{jq} or Eq.~\ref{jq-mf}.
\end{thebibliography}
\end{document}